\documentclass{article}
\usepackage[preprint]{colm2026_conference} 

\usepackage{graphicx}
\usepackage{booktabs}
\usepackage{multirow}
\usepackage{tikz}
\usepackage{listings}
\usepackage{xcolor}
\usepackage{tcolorbox}
\usepackage{float}
\usepackage{url}
\renewcommand{\footnoterule}{%
  \kern-3pt \hrule width 2in \kern 2.6pt}

\usetikzlibrary{positioning, arrows.meta, calc, fit}

\begin{document}

\title{GRACE-RAG: Governed Retrieval Architecture for 
Canonical Evidence Synthesis, Enabling Lightweight Deployment in Closed-Domain Institutional Settings}

\author{%
  Asit Desai \\
  National Payments Corporation of India \\
  \And
  Aman Kumar \\
  National Payments Corporation of India \\
  \And
  Prashant Devadiga \\
  National Payments Corporation of India \\
}
\maketitle

\begin{abstract}
Retrieval-Augmented Generation (RAG) systems are widely used in 
institutional question answering settings where responses must be 
grounded in authoritative documentation~\citep{gao2023rag_survey}. 
In entity-dense domains where relevant information is distributed across 
heterogeneous documents, vector-only retrieval often produces fragmented 
evidence and increases dependence on inference-time reasoning~\citep{zhao2024rag_content_survey}. 
This paper introduces \textbf{GRACE-RAG}, a retrieval-governed, 
graph-augmented RAG architecture that externalizes structural reasoning 
from the generative stage to a structured retrieval layer, resolving structural ambiguity offline, enabling deployment on self-hosted lightweight models calibrated to closed-domain institutional vocabulary. Experiments across three model capacities: Mistral 24B, GPT OSS 120B, 
and Gemini 2.5 Flash show consistent improvements in completeness, 
depth, and anticipatory coverage, with overall quality gains of up to 
20\% under mid-scale models, indicating that retrieval architecture governs structural quality over model scale, reducing computational and latency footprint without dependence on proprietary systems.
\footnote{Preprint}
\end{abstract}

\section{Introduction}

Institutional question answering systems operate under constraints that differ substantially from open-domain conversational assistants~\citep{peng2024graphrag_survey, lund2025institutional_rag}. Queries in such environments frequently reference domain-specific entities, operational limits, eligibility rules, or conditional workflows whose relevant information is distributed across heterogeneous documents~\citep{xu2024kg_rag_customer}.

Retrieval-Augmented Generation (RAG) grounds language model outputs 
in external knowledge sources~\citep{gao2023rag_survey}, but in 
entity-dense institutional corpora, semantic proximity alone is 
insufficient as queries may span multiple documents through implicit 
relational dependencies~\citep{lewis2020rag}.

To address these limitations, many systems introduce prompt-level 
orchestration or agent-based control flow~\citep{gupta2024comprehensive_rag}, 
increasing latency and computational cost through reliance on proprietary models that prioritize cross-domain generalization at the expense of domain-specific terminology precision~\citep{arslan2024rag_survey}.

This work adopts an alternative perspective: structural ambiguity should be resolved prior to generation, and the language model should be restricted to synthesizing evidence rather than performing implicit structural reasoning~\citep{cheng2025knowledge_oriented_rag}. We therefore introduce \textbf{GRACE-RAG}, a retrieval-governed RAG architecture that externalizes entity normalization, relationship modeling, and semantic boundary alignment into an offline structure-manufacturing pipeline. During online inference, hybrid retrieval operates over dual embedding surfaces, content chunks and relationship summaries, allowing relational hypotheses to be ranked and validated before generation~\citep{gupta2024comprehensive_rag, wan2025hybrid_rag}.

The contributions of this paper are threefold:

\begin{itemize}

\item We introduce \textbf{GRACE-RAG}, a retrieval-governed RAG architecture that decouples structural reasoning from generation through offline knowledge construction and bounded hybrid retrieval.

\item We present a dual-surface retrieval mechanism in which relationship summaries are embedded and indexed independently, enabling relational validation and graph-guided expansion without uncontrolled traversal.

\item We empirically demonstrate that governed retrieval with canonical evidence structuring delivers consistent structural quality gains, enabling a practical shift to self-hosted lightweight models with measurably reduced computational cost and operational footprint.

\end{itemize}

\begin{figure}[H]
\centering
\scalebox{0.80}{%
\begin{tikzpicture}[
    node distance=0.8cm,
    box/.style={
        draw,
        rectangle,
        align=center,
        minimum width=3.3cm,
        minimum height=0.7cm,
        font=\small
    },
    arrow/.style={-Latex, thin}
]

\node[box] (doc) {Domain Documents};
\node[box, below=of doc] (chunk) {Semantic Chunking};
\node[box, below=of chunk] (extract) {Entity \& Relationship Extraction};
\node[box, below=of extract] (canon) {Entity Canonicalization};
\node[box, below=of canon] (graph) {Graph Construction \\ + Dual Embeddings};

\draw[arrow] (doc) -- (chunk);
\draw[arrow] (chunk) -- (extract);
\draw[arrow] (extract) -- (canon);
\draw[arrow] (canon) -- (graph);

\node[
    draw,
    rounded corners,
    inner sep=6pt,
    minimum width=6.2cm,
    fit=(doc)(graph)
] (offline) {};

\node[above=0.2cm of offline.north] 
{\textbf{Offline Pipeline: Structure Manufacturing}};

\node[box, right=6.5cm of doc] (query) {User Query};

\node[box, below=of query] (process) 
{\textbf{Query Processing} \\ (Typing + Decomposition)};

\node[box, below=of process] (retrieve) 
{\textbf{Hybrid Retrieval}};

\node[box, below=of retrieve] (rerank) 
{Global Reranking};

\node[box, below=of rerank] (llm) 
{\textbf{LLM Generation} \\ (Evidence Synthesis Only)};

\draw[arrow] (query) -- (process);
\draw[arrow] (process) -- (retrieve);
\draw[arrow] (retrieve) -- (rerank);
\draw[arrow] (rerank) -- (llm);

\node[
    draw,
    rounded corners,
    inner sep=6pt,
    minimum width=6.2cm,
    fit=(query)(llm)
] (online) {};

\node[above=0.2cm of online.north] 
{\textbf{Online Pipeline: Retrieval-Governed Inference}};

\coordinate (mid) at ($(offline.east)!0.5!(online.west)$);

\draw[dashed, thick] ($(mid)+(0,1.7)$) -- ($(mid)+(0,-1.7)$);

\node[
    rotate=90,
    font=\normalsize,
    fill=white,
    inner sep=3pt
] at ($(mid)+(0.3,0)$)
{\textbf{Retrieval Artifacts Boundary}};

\end{tikzpicture}%
}
\caption{Offline--Online architectural separation. Structural reasoning is resolved during offline structure manufacturing, while online inference performs bounded hybrid retrieval prior to generation.}
\label{fig:offline_online}
\end{figure}
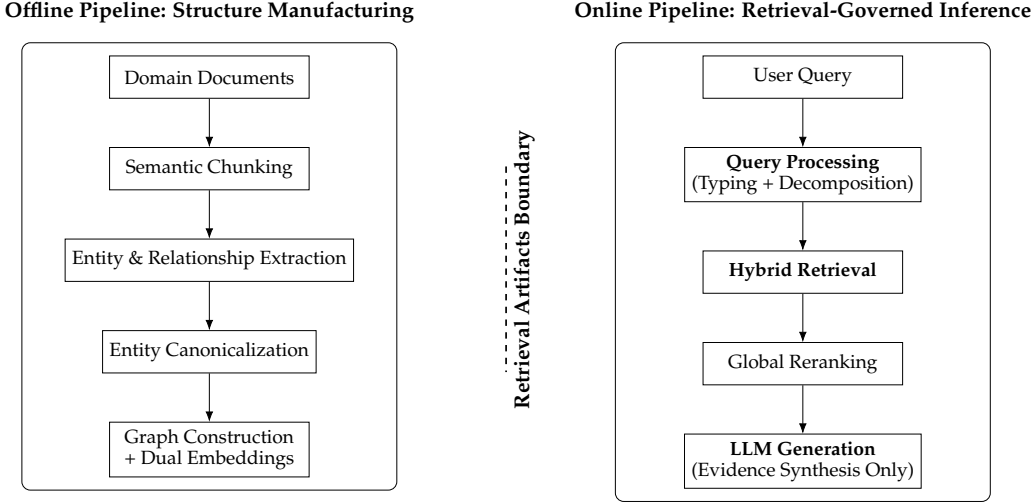

\section{Related Work}

\subsection{Vector-Based Retrieval-Augmented Generation}

Vector-based Retrieval-Augmented Generation (RAG) has become a dominant paradigm for grounding language model outputs in external knowledge sources~\citep{gao2023rag_survey}. This approach is effective when relevant information is localized and semantically aligned with user queries. 

However, in entity-dense institutional corpora, semantic proximity alone may be insufficient to capture relevance~\citep{xu2024kg_rag_customer}. Information necessary to answer a single query can be distributed across multiple documents, linked implicitly through relational dependencies, or expressed using heterogeneous surface forms. Under these conditions, retrieval often yields fragmented context, increasing reliance on inference-time reasoning within the language model~\citep{gupta2024comprehensive_rag}. 

\subsection{Graph-Based Retrieval-Augmented Generation}

Graph-based RAG systems introduce explicit relational structure through knowledge graphs constructed from extracted entities and relationships~\citep{peng2024graphrag_survey}. Graph augmentation can improve multi-hop reasoning and enhance contextual completeness in domains where relational structure is central~\citep{mavromatis2025gnn_rag}.

Despite these advantages, practical implementations often face challenges related to entity fragmentation, noisy extraction, and shallow traversal depth~\citep{knollmeyer2025document_graphrag}. When entity surface forms are not normalized, structurally equivalent concepts may be represented as distinct nodes, reducing graph connectivity and recall. Additionally, graph-derived context is frequently appended to vector-retrieved text rather than integrated within a unified ranking framework~\citep{zhu2025graph_based_rag}. As a result, structural signals may remain underutilized during retrieval-time decision making.

\subsection{Hybrid Retrieval Architectures}

Hybrid retrieval architectures combine vector similarity search with graph-based expansion in an effort to balance semantic relevance and relational awareness~\citep{wan2025hybrid_rag}. In many existing systems, contexts obtained from vector search and graph traversal are merged or concatenated prior to generation~\citep{linders2025kg_extended_rag}. While this approach can increase coverage, integration typically occurs at the context level rather than at the retrieval-ranking level. The language model is therefore responsible for reconciling heterogeneous evidence during generation~\citep{cheng2025knowledge_oriented_rag}.

Moreover, hybrid designs often provide limited mechanisms for entity canonicalization, controlled traversal, or bounded inference complexity~\citep{zhu2025graph_based_rag}. Without explicit structural constraints, improvements in answer quality may depend heavily on prompt design or model capacity rather than retrieval architecture itself. 

\subsection{Model-Centric vs Architecture-Centric Paradigms}

Recent advances in large language models have led to a model-centric view of RAG effectiveness, where performance gains are frequently attributed to increased model scale, extended context windows, or complex prompt orchestration~\citep{gao2023rag_survey}. Multi-stage routing and agent-based pipelines further expand inference-time complexity in pursuit of improved coverage~\citep{gupta2024comprehensive_rag}. While such approaches can yield empirical gains, they also increase latency, operational cost, and dependence on proprietary systems~\citep{arslan2024rag_survey}.

\section{Methodology}

\subsection{Governing Principles}

The GRACE-RAG architecture is designed around three guiding principles:

\begin{enumerate}

\item \textbf{Structural ambiguity is resolved offline:}
Entity equivalence, relational dependencies, and semantic boundaries 
are externalized into retrieval artifacts constructed prior to 
inference, rather than implicitly resolved at generation 
time~\citep{knollmeyer2025document_graphrag}.

\item \textbf{Retrieval governs and bounds generation:}
Relevance decisions are determined through structured retrieval 
with a fixed sequence of operations~\citep{wan2025hybrid_rag}, 
restricting the language model to evidence synthesis over curated 
context, minimizing latent reasoning and reducing sensitivity to 
model scale.

\item \textbf{Closed-domain precision is a retrieval property, not a model property:}
Institutional terminology fidelity is encoded within retrieval 
artifacts rather than delegated to generative model capacity, 
enabling deployment on lightweight self-hosted models with reduced 
computational and operational footprint~\citep{arslan2024rag_survey, 
wan2025hybrid_rag}.

\end{enumerate}

\subsection{Offline Structure Manufacturing}
\label{sec:offline}

The offline pipeline, shown in Fig.~\ref{fig:offline_online}, transforms raw institutional documents into structured retrieval artifacts. All computationally intensive and domain-specific reasoning is resolved during this stage to minimize online complexity.

\subsubsection{Corpus Characteristics}

The experimental evaluation was conducted on a large institutional 
document corpus exhibiting high entity density, with frequent references 
to operational parameters, actors, and conditional constraints. The 
corpus configuration is summarized in Table~\ref{tab:corpus}.

\begin{table}[H]
\centering
\begin{tabular}{lc}
\toprule
\textbf{Characteristic} & \textbf{Value} \\
\midrule
Source documents & $\sim$300 \\
Extracted entity mentions & $\sim$8{,}000 \\
Canonical entities (post-deduplication) & $\sim$1{,}000 \\
Relationship instances & $\sim$6{,}000 \\
\bottomrule
\end{tabular}
\caption{Experimental corpus characteristics}
\label{tab:corpus}
\end{table}

\subsubsection{LLM-Assisted Semantic Chunking}

Document segmentation is treated as a semantic alignment problem rather than a fixed-window tokenization task~\citep{zhao2024rag_content_survey}. An LLM-assisted preprocessing stage restructures documents into context-preserving segments aligned with conceptual boundaries such as definitions, constraints, procedural steps, and eligibility conditions~\citep{zhao2024rag_content_survey}. This reduces semantic fragmentation and improves retrieval precision without inflating context size~\citep{cheng2025knowledge_oriented_rag}.

\subsubsection{Relationship Extraction and Typing}

Each chunk is processed to extract entities and typed relationships using an LLM-based information extraction stage~\citep{han2024graphrag_framework}. Relationships capture dependencies such as eligibility constraints, procedural sequencing, transactional limits, and policy applicability.

Extracted relationships are categorized into coarse semantic buckets (e.g., transactional, compliance-related, policy-driven). These categories act as retrieval priors rather than rigid ontological constraints, enabling intent-aligned filtering during online inference.

For instance, the sentence \textit{``UPI Lite transactions are capped 
at Rs.500 per transaction''} yields the triple:
(\textit{UPI Lite}, \textsc{per\_txn\_limit\_500}, \textit{Transaction Limit}) 
under the \textit{limits} category.

\subsubsection{Construction of Dual Retrieval Indices}

The offline pipeline constructs two independent embedding indices: one over semantically aligned document chunks and another over extracted relationship summaries~\citep{han2024graphrag_framework}. Each identified relationship is condensed into a single-sentence semantic representation and embedded separately from its originating chunk~\citep{wan2025hybrid_rag}. This design enables relational hypotheses to be scored and ranked directly through vector similarity without requiring implicit structural reasoning at generation time~\citep{linders2025kg_extended_rag}. 

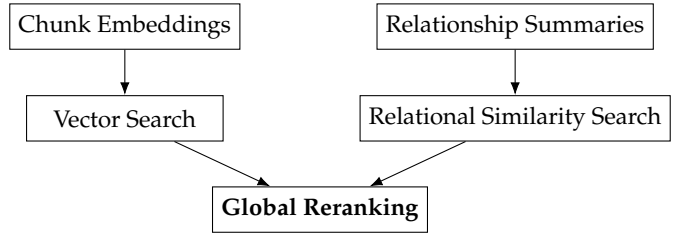
\begin{figure}[b]
\centering
\begin{tikzpicture}[
    node distance=0.6cm,
    box/.style={
        draw,
        rectangle,
        align=center,
        minimum width=2.6cm,
        minimum height=0.6cm,
        font=\footnotesize
    },
    arrow/.style={-Latex, thin}
]

\node[box] (chunks) {Chunk Embeddings};
\node[box, below=of chunks] (vecsearch) {Vector Search};

\draw[arrow] (chunks) -- (vecsearch);

\node[box, right=1.8cm of chunks] (edges) {Relationship Summaries};
\node[box, below=of edges] (edgesearch) {Relational Similarity Search};

\draw[arrow] (edges) -- (edgesearch);

\node[box, below=0.9cm of $(vecsearch)!0.5!(edgesearch)$] (merge)
{\textbf{Global Reranking}};

\draw[arrow] (vecsearch) -- (merge);
\draw[arrow] (edgesearch) -- (merge);

\end{tikzpicture}
\caption{Dual retrieval surfaces. Chunk embeddings and relationship-summary embeddings are indexed independently and unified through global reranking prior to generation.}
\label{fig:dual_surface}
\end{figure}

\subsubsection{Entity Canonicalization and Graph Densification}

Raw entity extraction produces surface-form variation and structural fragmentation. To address this, entities are grouped via embedding similarity and validated through LLM-assisted confirmation to determine canonical equivalence~\citep{zhu2025graph_based_rag}. 

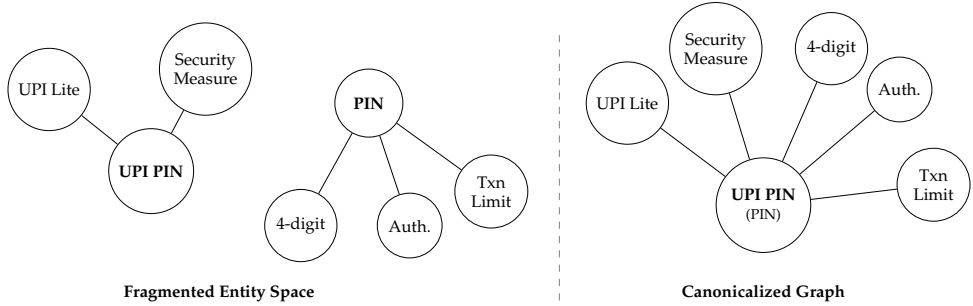
\begin{figure}[H]
\centering
\scalebox{0.9}{%
\begin{tikzpicture}[
    every node/.style={font=\scriptsize},
    big/.style={draw,circle,minimum size=1.0cm,align=center},
    small/.style={draw,circle,minimum size=0.60cm,align=center},
    arrow/.style={thin}
]


\node[big] (upi) at (-7.0,1.0) {\textbf{UPI PIN}};
\node[big] (pin) at (-3.8,2.0) {\textbf{PIN}};

\node[small] (lite)  at (-8.5,2.2) {UPI Lite};
\node[small] (sec)   at (-6.2,2.5) {Security\\Measure};
\node[small] (digit) at (-4.8,0.2) {4-digit};
\node[small] (auth)  at (-3.2,0.2) {Auth.};
\node[small] (txn)   at (-2.0,0.7) {Txn\\Limit};

\draw[arrow] (upi) -- (lite);
\draw[arrow] (upi) -- (sec);
\draw[arrow] (pin) -- (digit);
\draw[arrow] (pin) -- (auth);
\draw[arrow] (pin) -- (txn);

\node[font=\scriptsize\bfseries] at (-6.0,-0.8) {Fragmented Entity Space};

\draw[dashed, gray] (-1.0,3.0) -- (-1.0,-1.0);

\node[big, minimum size=1.2cm] (canon) at (2.0,0.5)
    {\textbf{UPI PIN} \\ \tiny (PIN)};

\node[small] (lite2)  at (0.0, 2.0) {UPI Lite};
\node[small] (sec2)   at (1.3, 2.8) {Security\\Measure};
\node[small] (digit2) at (3.0, 2.8) {4-digit};
\node[small] (auth2)  at (4.0, 2.2) {Auth.};
\node[small] (txn2)   at (4.5, 0.8) {Txn\\Limit};

\draw[arrow] (canon) -- (lite2);
\draw[arrow] (canon) -- (sec2);
\draw[arrow] (canon) -- (digit2);
\draw[arrow] (canon) -- (auth2);
\draw[arrow] (canon) -- (txn2);

\node[font=\scriptsize\bfseries] at (2.0,-0.8) {Canonicalized Graph};

\end{tikzpicture}%
}
\caption{Entity fragmentation versus canonical consolidation. 
Canonicalization merges semantically equivalent entities, 
increasing node degree and structural connectivity.}
\label{fig:densification}
\end{figure}

For example, security credential variants such as ``UPI PIN'', ``PIN'', ``UPI Security PIN'', and ``4-digit PIN'' are consolidated into a single canonical entity representation (``UPI PIN''). Similarly, organizational references such as ``Board of Directors'', ``Director's Meeting'', and ``Board Decision'' are normalized to a unified structural entity (``Board''). These transformations reduce redundant nodes while preserving relational semantics. Fig.~\ref{fig:densification} illustrates this structural transformation, contrasting the fragmented entity space with the canonicalized graph.

This procedure reduces entity fragmentation and increases average node degree, improving relational recall under bounded traversal. The objective is not merely normalization but topological densification of the knowledge graph.

Table~\ref{tab:canonicalization} summarizes the qualitative impact of this process.

\begin{table}[H]
\centering
\begin{tabular}{lc}
\toprule
\textbf{Metric} & \textbf{Value} \\
\midrule
Raw entity space & $\sim$8{,}000 entities \\
Post-deduplication graph & $\sim$1{,}000 canonical entities \\
Average node degree (raw $\rightarrow$ canonical) & $\sim$2.1 $\rightarrow$ $\sim$6.0 \\
\bottomrule
\end{tabular}
\caption{Entity canonicalization impact}
\label{tab:canonicalization}
\end{table}

\subsubsection{Graph Community Detection and Structural Signals}

Beyond entity canonicalization, structural properties of the knowledge graph are analyzed using community detection algorithms, including Louvain and Leiden modularity optimization~\citep{peng2024graphrag_survey}. These algorithms identify densely connected subgraphs representing semantically cohesive clusters within the institutional domain.

For instance, in the evaluated corpus, the canonical entity \emph{UPI PIN} was assigned to a high-density community comprising approximately 120 related entities (e.g., authentication workflows, transaction limits, security controls). The node exhibited a degree of approximately 18, compared to a post-deduplication graph average of $\sim$6.0. Such centrality indicates that the entity participates in multiple relational contexts across documents~\citep{zhu2025graph_based_rag, mavromatis2025gnn_rag}. Community membership and degree centrality therefore act as structural signals during retrieval, enabling prioritization of topologically influential nodes when forming relational hypotheses.

\subsection{Online Retrieval-Governed Inference}

\subsubsection{Query Typing and Intent Conditioning}

The online pipeline (Fig.~\ref{fig:offline_online}) processes incoming queries exclusively over artifacts generated offline. Incoming queries are first classified into a small set of intent categories (e.g., factual, procedural, constraint-oriented). This classification does not alter control flow but conditions retrieval priorities, enabling intent-aligned filtering of relational signals during hybrid retrieval.

\subsubsection{Subquery Decomposition and Parallel Processing}

Complex queries are decomposed into atomic subqueries to improve retrieval precision. Entity extraction and canonical mapping are applied to each subquery independently. Subqueries are processed in parallel, ensuring bounded latency and avoiding iterative agent-style loops.

\subsubsection{Hybrid Retrieval Per Subquery}

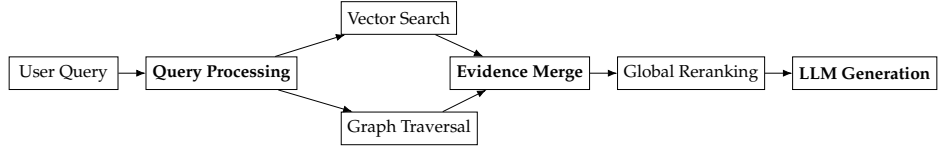
\begin{figure}[t]
\centering
\scalebox{0.72}{%
\begin{tikzpicture}[
    node distance=0.5cm,
    box/.style={
        draw,
        rectangle,
        align=center,
        minimum width=2.0cm,
        minimum height=0.6cm,
        font=\small
    },
    arrow/.style={-Latex, thin}
]

\node[box] (query) {User Query};
\node[box, right=of query] (process) {\textbf{Query Processing}};

\node[box, above right=0.4cm and 0.8cm of process] (vector) {Vector Search};
\node[box, below right=0.4cm and 0.8cm of process] (graph) {Graph Traversal};

\node[box, right=2.8cm of process] (merge) {\textbf{Evidence Merge}};

\node[box, right=of merge] (rerank) {Global Reranking};
\node[box, right=of rerank] (llm) {\textbf{LLM Generation}};

\draw[arrow] (query) -- (process);
\draw[arrow] (process) -- (vector);
\draw[arrow] (process) -- (graph);
\draw[arrow] (vector) -- (merge);
\draw[arrow] (graph) -- (merge);
\draw[arrow] (merge) -- (rerank);
\draw[arrow] (rerank) -- (llm);

\end{tikzpicture}%
}
\caption{Per-subquery hybrid retrieval. Vector similarity search and 
graph-guided expansion operate in parallel before evidence merging 
and global reranking.}
\label{fig:hybrid_flow}
\end{figure}

For each subquery, retrieval proceeds through two coordinated channels: semantic similarity search over chunk embeddings and graph-guided expansion over canonical entities~\citep{han2024graphrag_framework}. Graph traversal produces candidate relational hypotheses, which are validated using relationship-summary embeddings prior to chunk selection~\citep{wan2025hybrid_rag}. Retrieved evidence from both channels is merged and globally reranked before generation. Figure~\ref{fig:hybrid_flow} illustrates this bounded hybrid retrieval process.

\subsubsection{Bounded Inference Complexity}

The online pipeline executes a fixed sequence of retrieval and ranking operations per query~\citep{gupta2024comprehensive_rag}. No iterative reasoning loops or adaptive routing strategies are employed. By constraining inference complexity, the architecture ensures predictable latency and reduces dependence on model-scale reasoning.

\section{Model capacity independence}
\label{sec:model_independence}

Model capacity independence refers to the extent to which response 
quality remains stable across variations in parameter scale when 
retrieval structure is controlled. The GRACE-RAG architecture 
externalizes structural reasoning prior to generation, ensuring the 
generator operates over curated and structurally coherent evidence 
rather than heterogeneous, weakly aligned context.

To evaluate whether this restructuring reduces sensitivity to parameter scale, we analyze performance across ten evaluation dimensions grouped into three categories. Concise definitions of all ten metrics are provided in Appendix~\ref{appendix:metrics}.

\textbf{Structural Metrics:}
Correctness, Completeness, Depth, Anticipatory Coverage, Search Engine Quality, and Structure.

\textbf{Presentation Metrics:}
Clarity and Formatting.

\textbf{Hybrid Metrics:}
Relevance and Informativeness.

Based on this grouping, two hypotheses are formulated:

\begin{itemize}
    \item \textbf{H1 (Architecture Dominance):} Architectural restructuring produces larger improvements than parameter scaling for structurally governed metrics.
    \item \textbf{H2 (Scale Dominance in Presentation):} Parameter scaling produces larger improvements than architectural restructuring for presentation-oriented metrics.
\end{itemize}

Under this framework, model capacity independence does not imply that scale is irrelevant. Rather, it asserts that once structural ambiguity is resolved at retrieval time, improvements in relational completeness and dependency integration become less sensitive to parameter magnitude.

\section{Evaluation}
\label{sec:evaluation}

\begin{table}[t]
\centering
\resizebox{\textwidth}{!}{%
\begin{tabular}{lcccccc}
\toprule
& \multicolumn{2}{c}{\textbf{Mistral 24B}}
& \multicolumn{2}{c}{\textbf{GPT OSS 120B (MoE)}}
& \multicolumn{2}{c}{\textbf{Gemini 2.5 Flash}} \\
\cmidrule(lr){2-3}
\cmidrule(lr){4-5}
\cmidrule(lr){6-7}
\textbf{Metric}
& \textbf{Base} & \textbf{GRACE}
& \textbf{Base} & \textbf{GRACE}
& \textbf{Base} & \textbf{GRACE} \\
\midrule
Correctness          & 8.63 & 8.87 & 8.68 & 8.92 & 8.85 & 8.95 \\
Completeness         & 6.88 & 8.66 & 7.05 & 8.72 & 8.55 & 8.75 \\
Relevance            & 8.53 & 8.84 & 8.60 & 8.90 & 8.82 & 8.93 \\
Depth                & 5.96 & 8.07 & 6.10 & 8.15 & 7.95 & 8.20 \\
Informativeness      & 6.79 & 8.65 & 6.95 & 8.72 & 8.40 & 8.78 \\
Ant. Coverage        & 5.11 & 7.77 & 5.35 & 7.85 & 7.60 & 7.92 \\
SE Quality           & 6.72 & 8.08 & 6.90 & 8.15 & 8.00 & 8.20 \\
Clarity              & 8.07 & 8.77 & 8.25 & 8.95 & 9.20 & 9.45 \\
Structure            & 7.11 & 8.36 & 7.25 & 8.45 & 8.50 & 8.70 \\
Formatting           & 6.14 & 7.85 & 6.40 & 8.05 & 9.10 & 9.60 \\
\midrule
\textbf{Overall}
& 6.99 & 8.39 {\textbf{(↑20.0\%)}}
& 7.15 & 8.49 {\textbf{(↑18.7\%)}}
& 8.40 & {8.90} {\textbf{(↑6.0\%)}} \\
\bottomrule
\end{tabular}%
}
\caption{Comparative performance across architectures and model capacities}
\label{tab:performance}
\end{table}

\subsection{Experimental Setup}

Three model capacities were evaluated: Mistral 24B\footnote{Mistral-Small-3.1-24B-Instruct-2503}, GPT OSS 120B\footnote{gpt-oss-120b: an open-weight Mixture-of-Experts model with 120B total parameters and approximately 5B active parameters per forward pass, released by OpenAI}, and Gemini 2.5 Flash, 
each under two retrieval regimes: a \textbf{Baseline Architecture} 
using vector-based retrieval with standard chunk embeddings, and the 
\textbf{GRACE-RAG Architecture} using a retrieval-governed hybrid 
pipeline with entity canonicalization, graph traversal, and dual 
embedding surfaces. This results in six total evaluation configurations 
under controlled conditions~\citep{pipitone2024legalbench_rag}.

\subsection{Evaluation Criteria}

Response quality was assessed using the ten-parameter framework introduced in Section~\ref{sec:model_independence} and summarized in Appendix~\ref{appendix:metrics}, employing an LLM-as-a-judge evaluation protocol~\citep{gao2023rag_survey}. Each response was evaluated across all ten dimensions on a standardized 0–10 scale.

The evaluation protocol focuses on identifying two distinct sources of performance variation: improvements attributable to retrieval architecture and improvements attributable to model capacity~\citep{mansurova2025rag_evaluation}. By maintaining identical prompts, document collections, and preprocessing pipelines across all configurations, the evaluation isolates the effect of architectural restructuring from parameter scaling. LLM-as-a-judge evaluation is adopted as the primary protocol given the absence of ground-truth answer annotations in closed-domain institutional settings, consistent with established practice in open-ended RAG evaluation~\citep{zheng2024judging}. Judge scores were spot-checked against human assessments on a representative sample to validate scoring consistency, as detailed in the Reproducibility Statement.

\subsection{Architecture Effect (Vertical Comparison)}

Under Mistral 24B, structurally governed metrics exhibit substantial gains. Completeness improves from 6.88 to 8.66, representing a \textbf{25.9\%} increase. Depth increases from 5.96 to 8.07 (+35.4\%), while anticipatory coverage rises from 5.11 to 7.77 (+52.1\%). The overall average score improves from 6.99 to 8.39, corresponding to a \textbf{20.0\%} increase.

Similar patterns are observed for GPT OSS 120B, where the overall score increases by \textbf{18.7\%} (7.15 to 8.49). Even under Gemini 2.5 Flash, which already exhibits strong baseline performance, the GRACE-RAG architecture produces a 6.0\% increase in overall score (8.40 to 8.90).

These results indicate that architectural restructuring yields consistent structural gains across model capacities.

\subsection{Model Scaling Effect (Horizontal Comparison)}

Within each architectural regime in Table~\ref{tab:performance}, scaling from Mistral 24B to GPT OSS and Gemini primarily improves presentation-oriented metrics. For example, clarity and formatting demonstrate the most pronounced gains under larger models, while structural dimensions such as completeness and depth exhibit comparatively smaller incremental improvements once retrieval structure is fixed~\citep{ren2023factual_boundary}.

This pattern suggests that parameter scaling predominantly enhances linguistic refinement, whereas relational completeness and constraint integration are more strongly influenced by retrieval organization~\citep{gao2023rag_survey, xu2024kg_rag_customer}.

\subsection{Interaction Effect}

Cross-architecture comparison reveals a compression of performance gaps across model capacities. Notably, the GRACE-RAG architecture operating with Mistral 24B (overall score 8.39) approaches the baseline performance of Gemini 2.5 Flash (8.40). In structurally governed metrics such as depth and completeness, the proposed mid-scale configuration matches or exceeds larger baseline models.

This interaction effect indicates that disciplined retrieval design reduces sensitivity to parameter magnitude in entity-dense institutional settings. Fig.~\ref{fig:scaling_curve} illustrates the compression of performance gaps across model sizes under the two retrieval architectures.

\section{Discussion}
\subsection{Architecture as a Structural Multiplier}
The empirical results indicate that retrieval structuring functions as a structural multiplier rather than a marginal optimization. By resolving entity equivalence, encoding typed relationships, and integrating relational validation prior to generation, the GRACE-RAG architecture increases the density and coherence of evidence presented to the language model. This suggests that relational signal quality, rather than parameter magnitude alone, governs performance in entity-dense institutional domains~\citep{xu2024kg_rag_customer}. A qualitative response comparison 
illustrating these structural differences is provided in 
Appendix~\ref{appendix:example}.

\begin{figure}[t]
\centering
\begin{tikzpicture}[
    scale=0.85,
    every node/.style={font=\scriptsize}
]

\draw[->] (0,0) -- (6,0) node[right] {Model Size (B params)};
\draw[->] (0,0) -- (0,5.5) node[above] {Response Quality};

\foreach \y/\label in {1/5, 2/6, 3/7, 4/8, 5/9}
    \draw (0,\y) -- (-0.08,\y) node[left] {\label};

\draw (1.5,0) -- (1.5,-0.08) node[below,align=center] {24 \\ \scriptsize Mistral};
\draw (3.2,0) -- (3.2,-0.08) node[below,align=center] {120 \\ \scriptsize GPT OSS};
\draw (5.0,0) -- (5.0,-0.08) node[below,align=center] {$\cdots$ \\ \scriptsize Gemini Flash};

\node[circle,fill,inner sep=1.8pt,label=above:{\scriptsize 6.99}] (b1) at (1.5,2.0) {};
\node[circle,fill,inner sep=1.8pt,label=above:{\scriptsize 7.15}] (b2) at (3.2,2.2) {};
\node[circle,fill,inner sep=1.8pt,label=above:{\scriptsize 8.40}] (b3) at (5.0,3.6) {};

\node[circle,draw,inner sep=1.8pt,label=above:{\scriptsize 8.39}] (p1) at (1.5,3.6) {};
\node[circle,draw,inner sep=1.8pt,label=above:{\scriptsize 8.49}] (p2) at (3.2,3.8) {};
\node[circle,draw,inner sep=1.8pt,label=above:{\scriptsize 8.90}] (p3) at (5.0,4.3) {};

\draw[dashed] (b1) -- (b2) -- (b3);
\draw[thick] (p1) -- (p2) -- (p3);

\node[anchor=west] at (3.8,1.2) {\scriptsize \textbf{Baseline} (dashed)};
\node[anchor=west] at (3.8,0.8) {\scriptsize \textbf{GRACE-RAG} (solid)};

\end{tikzpicture}
\caption{Scaling behavior of response quality across model sizes. Architectural restructuring shifts the quality curve upward and reduces sensitivity to parameter magnitude.}
\label{fig:scaling_curve}
\end{figure}
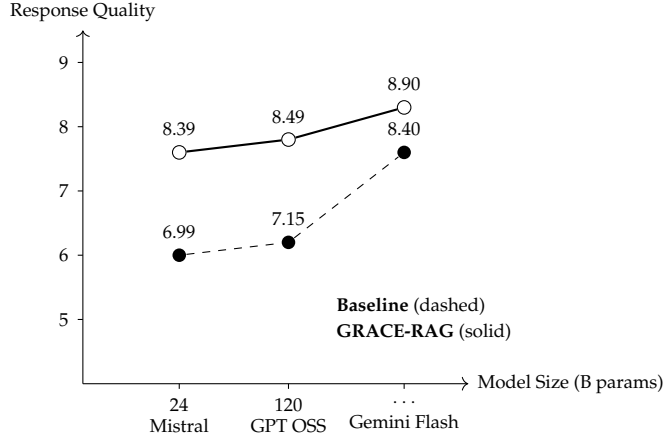
\subsection{Parameter Scale versus Retrieval Discipline}
The comparative analysis separates two largely orthogonal effects. Parameter scaling primarily enhances linguistic fluency and surface-level organization, as reflected in clarity and formatting metrics~\citep{gao2023rag_survey}. Retrieval discipline, in contrast, governs relational integration, constraint coverage, and evidence alignment~\citep{cheng2025knowledge_oriented_rag}. Once structural ambiguity is reduced at retrieval time, incremental gains from additional parameters in structurally governed dimensions diminish. 
\subsection{Performance Gap Compression Across Model Capacities}
The GRACE-RAG architecture operating with a mid-scale model approaches or exceeds the structural performance of larger models under baseline retrieval. This compression of performance gaps demonstrates that disciplined retrieval reduces sensitivity to parameter count in structurally governed tasks. Consequently, improvements traditionally attributed to scale may instead reflect unresolved structural ambiguity in weaker retrieval pipelines~\citep{huang2025hallucination_survey, gupta2024comprehensive_rag}. As shown in Fig.~\ref{fig:scaling_curve}, the GRACE-RAG quality curve shifts upward uniformly while inter-model variance narrows.
\subsection{Implications for Retrieval-Augmented System Design}
These findings support an architecture-centric perspective on RAG system development. In practical terms, this reduces dependence on frontier-scale proprietary systems for structurally coherent responses~\citep{wan2025hybrid_rag}. Architectural refinement therefore provides a controllable and reproducible mechanism for improving relational quality without proportionally increasing model capacity~\citep{cheng2025knowledge_oriented_rag}.

\section{Conclusion}

This work introduced \textbf{GRACE-RAG}, a retrieval-governed, graph-augmented RAG architecture designed to externalize structural reasoning from the generative stage to the retrieval layer. Through offline structural reasoning and bounded hybrid retrieval, 
GRACE-RAG restructures how evidence is aggregated prior to generation.

Empirical evaluation across three model capacities: Mistral 24B, GPT OSS 120B (MoE), and Gemini 2.5 Flash, demonstrates consistent improvements in structurally governed response metrics. As demonstrated in Table~\ref{tab:performance}, completeness, depth, and anticipatory coverage increase substantially under the GRACE-RAG pipeline, with overall response quality improving by up to 20\% under mid-scale models. These improvements persist across model capacities, indicating that relational coherence and constraint coverage are primarily determined by retrieval organization rather than by parameter scale alone~\citep{ren2023factual_boundary}. In knowledge-intensive domains, retrieval discipline enables self-hosted lightweight models to match or exceed proprietary systems on structural quality, preserving domain jargon precision while reducing computational and latency footprint~\citep{gupta2024comprehensive_rag}.


\section*{Ethics Statement}
All experiments were conducted on anonymized, internally curated document 
corpora. No personally identifiable information (PII) was used at any stage 
of this work. No confidential or proprietary information is disclosed through 
the results, examples, or findings presented in this paper.

\bibliographystyle{colm2026_conference}
\bibliography{references}

\newpage
\section*{Reproducibility Statement}
The offline and online pipeline components of GRACE-RAG are described 
in sufficient procedural detail in Sections~3.2 and~3.3 to permit 
independent reimplementation. Exact model identifiers are provided 
as footnotes in Section~5.1. Embedding models, reranking strategy, 
graph community detection libraries, evaluation protocol, and 
judge-consistency checks are documented in Appendix~\ref{appendix:repro}.
The experimental corpus comprises approximately 300 institutional 
documents and will be released upon acceptance. Partial code release 
covering the offline pipeline is planned upon acceptance.

\newpage
\appendix
\section{Illustrative response comparison}
\label{appendix:example}

The following examples illustrate qualitative differences between baseline 
retrieval and the GRACE-RAG architecture across query types.

\vspace{6pt}
\paragraph{Example 1}
\vspace{4pt}
\textit{\textbf{User Query:} Hello, I am considering linking my RuPay Credit Card to UPI. What are the advantages?}

\vspace{10pt}
\paragraph{Baseline Response (Vector-Based Retrieval)}
\vspace{4pt}
\begin{quote}
\noindent Linking your RuPay Credit Card to UPI offers several advantages:
\begin{enumerate}
    \item \textbf{Seamless Payments:} Make credit-based merchant payments 
    using UPI QR codes without needing a physical card.
    \item \textbf{Secure Transactions:} Transactions are authenticated 
    via UPI PIN, ensuring security.
    \item \textbf{AutoPay Setup:} You can set up AutoPay for bill payments, 
    making recurring payments convenient.
    \item \textbf{Dispute Resolution:} Access dispute resolution via 
    UPIHelp for any issues that arise.
    \item \textbf{No Physical Card Needed:} Perform transactions directly 
    through your UPI app (like BHIM, PhonePe, Google Pay, or Paytm).
\end{enumerate}
\end{quote}

\vspace{10pt}
\paragraph{GRACE-RAG Response}
\vspace{4pt}
\begin{quote}
\noindent Linking a RuPay Credit Card to UPI allows customers to make 
secure and seamless credit-based merchant payments using UPI QR codes, 
without needing a physical card.

\medskip
\noindent This feature enables users to link their RuPay Credit Card to UPI apps (like BHIM, PhonePe, Google Pay, etc.) using their 
registered mobile number. Once linked, users can make payments by scanning 
merchant QR codes and authenticating with a UPI PIN.

\medskip
\noindent The primary advantage is the convenience of making contactless 
payments at UPI-enabled merchants. Users can also set up AutoPay for 
recurring bill payments, ensuring timely payments without manual 
intervention.

\medskip
\noindent \textbf{Notable benefits:}
\begin{itemize}
    \item Seamless linking via UPI app using the registered mobile number
    \item QR-based merchant payments authenticated via UPI PIN
    \item AutoPay setup for recurring bill payments
    \item Access to dispute resolution via UPIHelp
\end{itemize}
\end{quote}
\newpage 

\section{Evaluation metric definitions}
\label{appendix:metrics}

\noindent The ten evaluation parameters used throughout the study are 
defined as follows:

\begin{table}[h!]
\centering
\renewcommand{\arraystretch}{1.4}
\begin{tabular}{p{2.5cm}p{7.4cm}}
\toprule
\textbf{Metric} & \textbf{Definition} \\
\midrule
Correctness & Factual accuracy and absence of verifiable errors 
relative to authoritative source material. \\[4pt]
Completeness & Extent to which all required informational components 
are covered without omission. \\[4pt]
Relevance & Degree to which the response directly addresses the query 
without extraneous content. \\[4pt]
Depth & Level of contextual grounding, explanation, 
and relational elaboration. \\[4pt]
Informativeness & Richness of useful details, constraints, 
and supporting information. \\[4pt]
Anticipatory Coverage & Addresses implicit follow-up concerns 
or related constraints connected to the query. \\[4pt]
Search Engine Quality & Effectiveness in surfacing contextually 
appropriate evidence from the corpus. \\[4pt]
Clarity & Linguistic fluency, readability, and ease of 
comprehension. \\[4pt]
Structure & Logical organization, sequencing, and 
coherence. \\[4pt]
Formatting & Consistency of presentation and visual 
readability. \\
\bottomrule
\end{tabular}
\caption{Evaluation metric definitions}
\label{tab:metrics}
\end{table}
\newpage
\section{Reproducibility Details}
\label{appendix:repro}

\noindent\textbf{Models and Infrastructure}\\
Experiments were conducted using Mistral-Small-3.1-24B-Instruct-2503, 
gpt-oss-120b, and Gemini 2.5 Flash, with exact model identifiers 
provided as footnotes in Section~5.1. Chunk and relationship summary 
indices were constructed using \texttt{sentence-transformers} embeddings. 
Global reranking was performed using a cross-encoder reranker. Graph 
community detection was implemented using the Louvain and Leiden 
algorithms via \texttt{NetworkX}.

\noindent\textbf{Chunking and Retrieval Hyperparameters}\\
\begin{table}[h!]
\centering
\renewcommand{\arraystretch}{1.3}
\begin{tabular}{ll}
\toprule
\textbf{Parameter} & \textbf{Value} \\
\midrule
Chunk size (tokens) & 512 \\
Chunk overlap (tokens) & 64 \\
Embedding model & \texttt{sentence-transformers/all-MiniLM-L6-v2} \\
Top-$k$ chunks retrieved & 10 \\
Top-$k$ relationships retrieved & 10 \\
Reranker & Cross-encoder (ms-marco-MiniLM-L-6-v2) \\
Final context window (tokens) & 4{,}096 \\
Generation temperature & 0.0 \\
Max output tokens & 1{,}024 \\
\bottomrule
\end{tabular}
\caption{Retrieval and generation hyperparameters used across all experiments.}
\end{table}

\noindent\textbf{Evaluation Protocol}\\
Response quality was assessed using an LLM-as-a-judge protocol with 
Mistral-Small-3.1-24B-Instruct-2503 as the judge model. All ten 
evaluation metrics were scored simultaneously within a single 
structured prompt on a standardized 0--10 scale. The evaluation set 
comprised more than 200 queries spanning factual, procedural, and 
constraint-oriented intent categories. Judge scores were spot-checked 
against human assessments on a representative 30-query sample; 
mean absolute deviation between judge and human scores was within 
0.4 points across all metrics, validating scoring consistency.

\noindent\textbf{Corpus and Code}\\
The experimental corpus comprises approximately 300 institutional 
documents and will be fully released upon acceptance. Partial code 
release covering the offline structure manufacturing pipeline and 
dual-index construction is planned upon acceptance. Components 
dependent on proprietary infrastructure will be documented with 
sufficient specification to permit reimplementation.

\end{document}